\documentclass[reprint,superscriptaddress,amsmath,amssymb,aps,pra,longbibliography]{revtex4-1}

\usepackage{graphicx} 
\usepackage{epstopdf}
\usepackage{caption}
\usepackage{subcaption}

\usepackage{dcolumn} 
\usepackage{bm} 
\usepackage{hyperref} 
\usepackage{color}
\usepackage{gensymb}

\usepackage[T1]{fontenc}
\usepackage{textcomp}

\begin{document}

\preprint{APS/123-QED}

\title{Active control of laser wavefronts in atom interferometers}

\author{A. Trimeche}
\author{M. Langlois}
\author{S. Merlet}
\author{F. Pereira Dos Santos}
\email{franck.pereira@obspm.fr}
\affiliation{LNE-SYRTE, Observatoire de Paris, PSL Research University, CNRS, Sorbonne Universit\'es, UPMC Univ. Paris 06, \\ 61 Avenue de l'Observatoire, 75014 Paris, France}

\date{\today}

\begin{abstract}
Wavefront aberrations are identified as a major limitation in quantum sensors. They are today the main contribution in the uncertainty budget of best cold atom interferometers based on two-photon laser beam splitters, and constitute an important limit for their long-term stability, impeding these instruments from reaching their full potential. Moreover, they will also remain a major obstacle in future experiments based on large momentum beam splitters. In this article, we tackle this issue by using a deformable mirror to control actively the laser wavefronts in atom interferometry. In particular, we demonstrate in an experimental proof of principle the efficient correction of wavefront aberrations in an atomic gravimeter. 
\begin{description}
\item[PACS numbers] 04.80.Nn, 03.75.Dg, 42.55.Ye, 42.15.Fr
\end{description}
\end{abstract}

\maketitle

\section{Introduction}

\begin{figure}[!ht]
	\centering
	\includegraphics[scale=0.55]{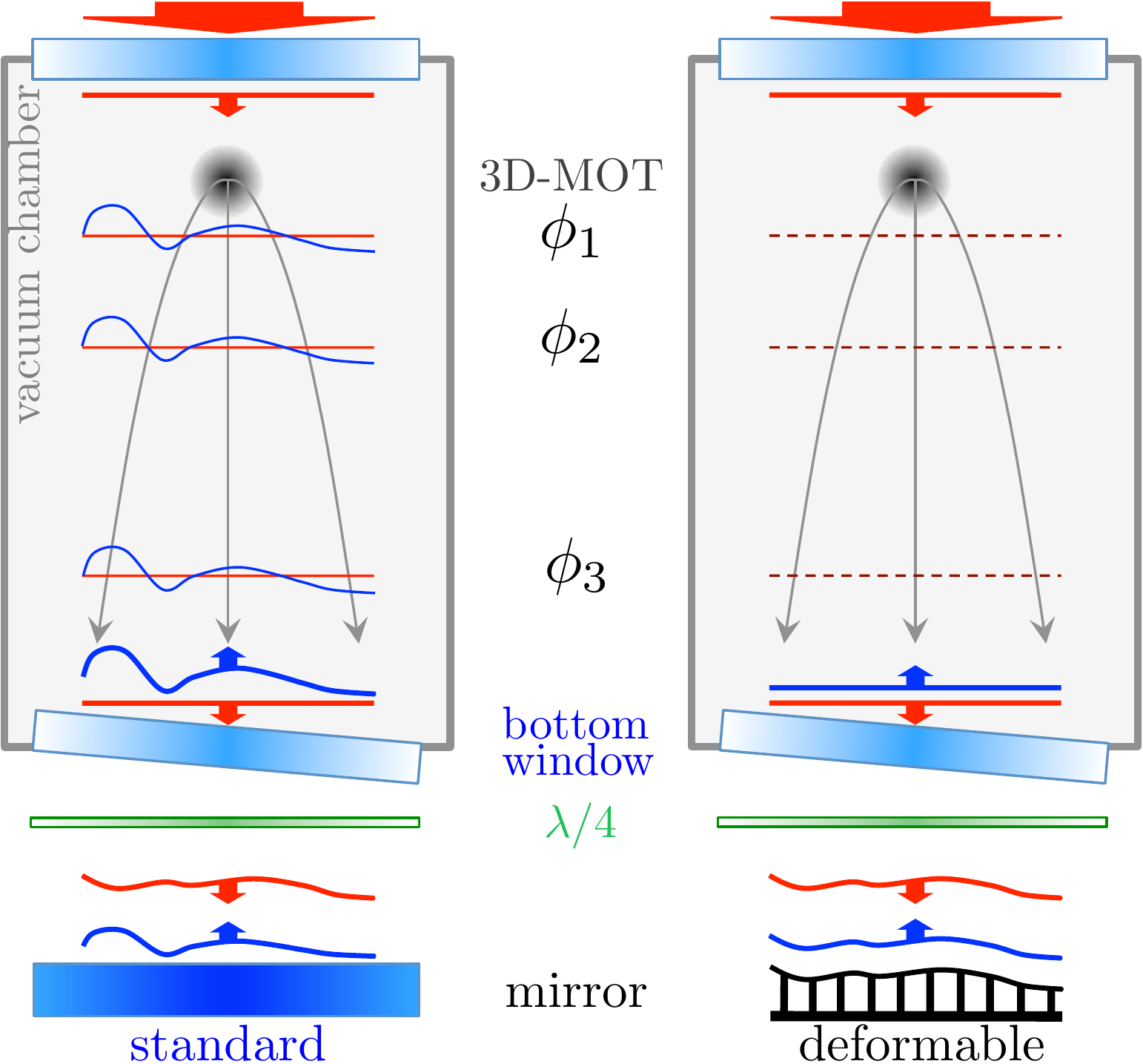}
	\caption{Laser wavefronts propagation. The laser beam enters the vacuum chamber from the top and exits through the bottom window. The descending wavefront is taken as flat (red). After being reflected by a standard mirror (left) or a deformable mirror (right), it re-enters the vacuum chamber (blue). Left: the ascending wavefront gets distorted by the aberrations of the bottom window, $\lambda$/4 plate and standard mirror. The laser phase difference then depends on the transverse position. It gets sampled differently at the three pulses depending on the ballistic trajectories of the atoms, which leads to a bias. Right: the ascending wavefront is corrected by properly shaping the deformable mirror. This leads to uniform laser phase differences and no bias.}
	\label{fig:Setup and principle}
\end{figure}

Inertial sensors based on atom interferometry \cite{Revue:09}, such as gravimeters and gradiometers \cite{Hu:13,Gillot:14,freier:16,Kasevich:02,Sorrentino:14} or gyroscopes \cite{Kasevich:00,Dutta:16}, are subject today to intense developments, owing to their large range of applications, in geophysics, navigation, space science and high precision measurements in fundamental physics \cite{Kasevich:07,Rosi:14,Bouchendira:11}. In light-pulse atom interferometers \cite{Borde:89}, the final phase shift depends on the acceleration and the rotation of the experimental setup with respect to the inertial reference frame defined by the atoms in free fall. The inertial force is then derived from the measurement of the relative displacement of these atoms compared to the lasers' equiphases. Distortions of these equiphases thus induce parasitic phase shifts which bias the measurement. This effect is linked to the residual ballistic motion of the atoms in the laser beam profile during their free fall as displayed on the left of Fig. \ref{fig:Setup and principle}. Wavefront aberrations are identified and measured on atom interferometers \cite{Louchet:10,Schkolnik:15,Zhou:16} as the major source of bias uncertainty and long-term instability in the best light-pulse atom interferometers used as inertial sensors, such as high precision gravimeters \cite{freier:16,Gillot:14} and gyroscopes \cite{Gauguet:09,Rasel:14,Dutta:16}. This is also true for next generation experiments, such as those based on large momentum beam splitters \cite{Kasevich:15}, as well as in future space projects \cite{Altschul:15}.

The influence of wavefront aberrations can in principle be limited, if not suppressed, by performing atom interferometry inside a cavity, such as in \cite{Muller:15}, which allows for spatial mode selection and filtering. Yet, the requirement of operating the interferometer with large laser waists, of order of a cm radius size, in a compact cavity puts severe constraints on the realization and alignment for stable operation and for avoiding the coupling of unwanted transverse modes, which otherwise induce large wavefront aberrations \cite{Riou:17}.

In astronomy, wavefront distortions and their fluctuations due to atmospheric turbulence also impose severe limits to the resolution of large area telescopes. To overcome this problem, deformable mirrors (DM) have been proposed \cite{Babcock:53} and developed \cite{Merkle:89} for efficient real time correction of wavefront aberrations. They are based on different technologies such as 9-actuator deformable electrostatic membrane using continuous voltage distribution \cite{Bonora:11}, 35-actuator bimorph deformable mirror composed of two disks of lead magnesium niobate \cite{Horsley:07}, thin polymer membrane with permanent magnets and microcoils \cite{Cugat:01}. DM are already used to correct wavefront aberrations of laser beams, potentially in closed loop \cite{Poyneer:06}, and for instance with thermally deformable mirror \cite{Kasprzack:13}, in various fields such as ophthalmology, optical beams interferometry and femtosecond pulse shaping. DM can also be used for tailoring the shape of the cavity eigenmodes \cite{Peter:06}, and thus selecting the coupled transverse modes. Last, it enables to generate flat-top laser beams \cite{Tarallo:07}, which are of interest for light pulse atom interferometry. 

Here, a DM is used for the first time to control the laser wavefront in an atom interferometer. We demonstrate its ability and efficiency to correct the wavefront aberrations in a proof of principle experiment realized with an atomic gravimeter.

\section{Description of the experiment}

The sensor head of the gravimeter is described in \cite{LeGouet:08}. The laser system, which is realized using two extended cavity laser diodes, and a typical measurement sequence are detailed in \cite{Merlet:14}. In this compact experimental setup, atoms are loaded directly from a background $^{87}Rb$ vapor, trapped in a three-dimensional magneto-optical trap (MOT), and further cooled down to 2~$\mu$K before being dropped in free fall by switching off the cooling lasers. The interferometer is obtained by pulsing counter-propagating laser beams in the vertical direction. Two co-propagating vertical laser beams, of wavevectors $\vec{k}_{1}$ and $\vec{k}_{2}$, are first overlapped and delivered to the atoms through a single collimator. The counter-propagating beams are obtained by reflection on a mirror. Due to the Doppler shift induced by the free fall of the atoms, only two counter-propagating beams will drive the stimulated Raman transitions according to the two-photon resonance condition. A three Raman pulse sequence $\frac{\pi}{2} - \pi - \frac{\pi}{2}$ allows to split, deflect, and recombine the atomic wave packets, thus realizing a Mach Zehnder type interferometer. With this geometry, the atomic phase-shift at the output of the interferometer is given by \cite{chu:91}: $\Delta\Phi=\phi_{1}-2\phi_{2}+\phi_{3}$, where $\phi_{\mathrm{i}}$ is the phase difference between the two Raman lasers, at the position $\vec{z}_{\mathrm{i}}$ of the center of mass of the wavepacket, at the time of the i-th Raman pulse. For ideal plane wavefront, $\phi_{i}^{p}=\vec{k}_{\mathrm{eff}}\vec{z}_{\mathrm{i}}$ which leads to $\Delta\Phi^{p}=-\vec{k}_{\mathrm{eff}}\vec{g}T^{2}$ where $\vec{k}_{\mathrm{eff}}=\vec{k}_{1}-\vec{k}_{2}$ is the effective wave-vector, $\vec{g}$ the acceleration of the Earth gravity, and $T$ the free-evolution time between two consecutive Raman pulses. Such atomic accelerometers are thus sensitive to the relative acceleration between the free-falling atoms and the retro-reflecting mirror, which sets the phase reference for the Raman lasers. Any deviation of the phase regarding to $\phi_{\mathrm{i}}^{p}$ might lead to a bias on the gravity measurement due to the expansion of the atomic cloud across the lasers wavefronts (see Fig. \ref{fig:Setup and principle}).

\section{Characterization of the mirror}

\begin{figure}[h]
	\centering
	\begin{subfigure}[b]{0.23\textwidth}
		\includegraphics[width=\textwidth]{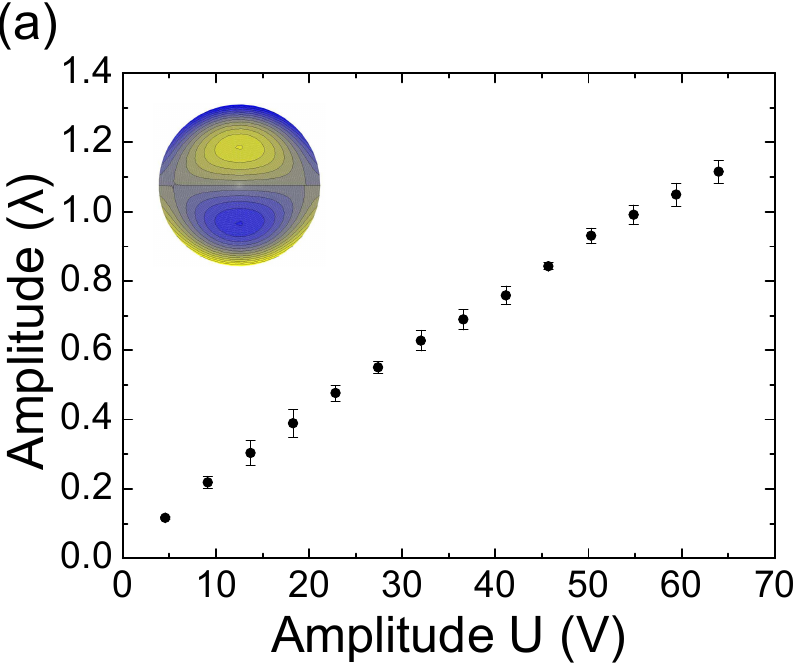}
	\end{subfigure}
	\begin{subfigure}[b]{0.23\textwidth}
		\includegraphics[width=\textwidth]{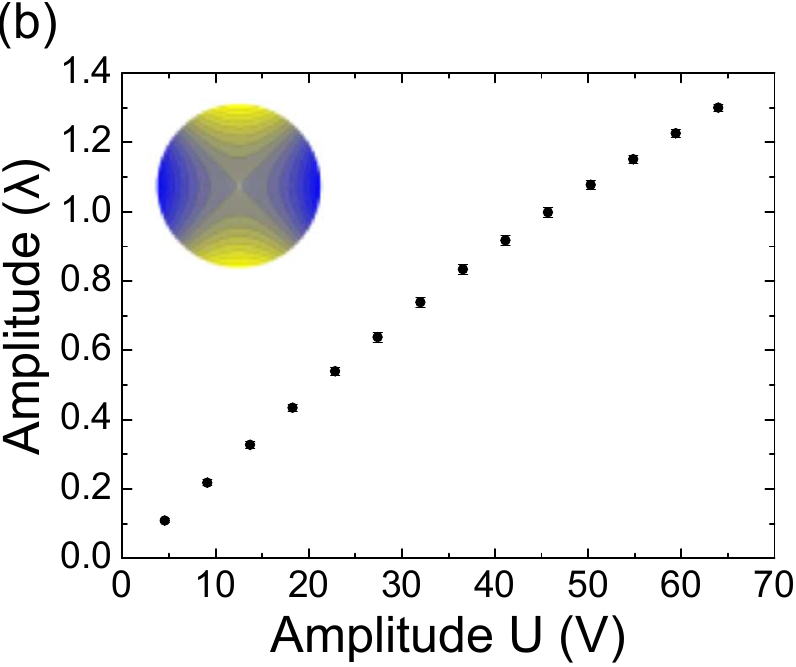}
	\end{subfigure}
	\caption{Amplitude of the DM deformation as a function of the amplitude $U$ of the applied voltage for (a) a coma 90$\degree$ deformation, and (b) an astigmatism 0$\degree$ deformation.}
	\label{fig:DMresponse}
\end{figure}

In our setup, the retro-reflecting mirror (and an additional quarter-wave plate) are placed outside the vacuum chamber as shown in Fig. \ref{fig:Setup and principle}. Formerly, as described in \cite{Merlet:14}, we have used a standard dielectric mirror and obtained at best a sensitivity of 60~$\mu$Gal in 1~s measurement time (1~$\mu$Gal=$10^{-8}$~ms$^{-2}$). For this study, the mirror was replaced by a Kilo-C-DM MEMS Deformable Mirror from Boston Micromachines Corporation with a 9.9~mm diameter of active circular surface. This DM uses 952 micro-actuators, with a pitch of 300~$\mu$m, and a maximum stroke of 1.8~$\mu$m for an applied voltage of 195~V. The DM surface is a continuous gold coated membrane, with a specified flatness of 11~nm RMS. A homemade software program allows to control the DM surface shape, by varying the amplitude of the first 64 Zernike polynomials \cite{Wang:80} which are conventionally used as a basis to decompose wavefront aberrations. The default setting of the DM is the Flat Map (FM) configuration, which is calibrated by the constructor in order to make the mirror plane, with optimized voltages for each actuator (around 80~V). This calibration is performed so as to minimize the RMS error, and the corresponding measurement performed with a wavefront sensor was provided. From this measurement, we calculate a flatness of 6.47~nm of RMS (10.81~nm RMS considering the DM rectangular edge), and 28.02~nm of Peak-to-Valley dominated by a residual curvature. This is comparable to the flatness of best high quality commercially available dielectric mirrors.

To characterise the response of the DM with respect to the applied voltage, we measured the wavefront deformation of a laser beam reflected by the DM using a Shack-Hartmann sensor \cite{Platt:01} (SH), a \textit{HASO} marketed by the company Imagine Optic. We deform the mirror by applying on each actuator $i$ a voltage $V(i)=V_{FM}(i)+U.Z(i)$, where $V_{FM}(i)$ is the setting of the FM configuration, $Z(i)$ is a given Zernike polynomial evaluated at the pixel $i$, and $U$ is the corresponding amplitude. We then performed differential measurements, subtracting from the deformed wavefront signal the reference FM wavefront. The measurement was performed on several aberrations, corresponding to the lowest order Zernike polynomials, which are expected to be dominant in our experiment. To illustrate these measurements, we display in figure 2 the response of the DM to the amplitude $U$ of the applied voltage for a coma 90$\degree$ deformation (Fig. \ref{fig:DMresponse} (a)) and for an astigmatism 0$\degree$ deformation (Fig. \ref{fig:DMresponse} (b)). For weak amplitudes $U$ (below about 20~V), we find a linear behavior of the actuators motion. Non-linearities at higher amplitudes make the DM response decrease. We measured an amplitude of 0.025(1)$\lambda$ per Volt added to the FM voltages, which is twice the surface deformations because of the reflection onto the DM. This is in perfect agreement with the constructor calibrations. In addition, we found the standard deviation of measurements repeated over several days to be lower than $\lambda/125$, limited by the SH repeatability, confirming the DM long-term stability in open loop \cite{Morzinski:06}. 

\begin{table}[h]
\centering
\caption{Interferometer phase-shifts due to different aberration orders of the retroreflecting mirror. The phase shifts are averaged over the velocity distribution for initial positions $x_{0}$ and $y_{0}$ of the atomic cloud. $f(R,x_{0},y_{0},t_{1},T,\sigma_{\nu})=4(x_{0}^{2}+y_{0}^{2}+\sigma_{\nu}^{2}(6t_{1}^{2}+12t_{1}T+7T^{2}))-R^{2}$ where $R$ is the mirror radius, $t_{1}$ is the delay of the first Raman pulse with respect to the release time of the atoms, and $\sigma_{\nu}$ is the initial velocity dispersion of the atomic cloud.}
\begin{tabular}{cc}
\hline
Zernike polynomial $\left( \mathrm{Z}_{n}^{m}\right)$ & $\Delta\Phi$ \\
\hline
Piston $\left( \mathrm{Z}_{0}^{0}\right)$, Tilts $\left( \mathrm{Z}_{1}^{\pm 1}\right)$ & 0 \\
Focus $\left( \mathrm{Z}_{2}^{0}\right)$ & $8k_{\mathrm{eff}}T^{2}\sigma_{\nu}^{2}/R^{2}$ \\
Astigmatisms $\left( \mathrm{Z}_{2}^{\pm 2}\right)$ & 0 \\
Comas 0$\degree$ $\left( \mathrm{Z}_{3}^{-1}\right)$ & $24k_{\mathrm{eff}}T^{2}x_{0}\sigma_{\nu}^{2}/R^{3}$ \\
Coma 90$\degree$ $\left( \mathrm{Z}_{3}^{1}\right)$ & $24k_{\mathrm{eff}}T^{2}y_{0}\sigma_{\nu}^{2}/R^{3}$ \\
Spherical ab. $\left( \mathrm{Z}_{4}^{0}\right)$ & $24k_{\mathrm{eff}}T^{2}\sigma_{\nu}^{2}f(R,x_{0},y_{0},t_{1},T,\sigma_{\nu})/R^{4} $ \\
\hline
\end{tabular}
\label{tab:PhaseShiftAberration}
\end{table}

The parasitic phase shifts induced by the wavefront aberrations of the laser beams, result from the convolution between the distribution of atomic trajectories and the Raman beam wavefronts, and consequently depend on many experimental parameters such us the temperature, the initial position and velocity distribution of the atomic cloud, the shape of the Raman beams, etc... Table \ref{tab:PhaseShiftAberration} lists the expected phase shifts at the output of the interferometer, induced by the most common aberrations, which correspond to some of the first Zernike polynomials. These phase shift formulas are derived for Raman beams with infinite size and homogeneous intensity profile, and for a point source atomic cloud in ballistic expansion.

The focus gives an interferometer phase shift independent of the initial positions of the atoms. On the contrary, the shifts due to comas depend linearly on the initial positions. It is thus in principle zero when the atomic distribution is centered on the mirror. We actually use this linear dependence to center the atomic cloud on the mirror or/and the mirror on the atomic cloud (see below). As for the astigmatism, it is zero and thus independent of the initial positions of the atoms. However, this is related to averaging the effects of opposite curvatures along orthogonal directions and assumes a radial isotropy. This no longer holds if the velocity distribution (of the detected atoms) is not isotropic, which can be induced by spatial inhomogeneities of the detection. For instance, with gaussian velocity distributions, eventually different along two orthogonal directions, we obtain: $\Delta\Phi=2k_{eff}T^{2}(\sigma_{\nu_{x}}^{2}-\sigma_{\nu_{y}}^{2})/R^{2}$, where $\sigma_{\nu_{x,y}}$ are the projection of the initial velocity dispersion of the atomic cloud.

Remarkably, most of the first modes of Table \ref{tab:PhaseShiftAberration} depend on $T^{2}\sigma_{\nu}^{2}$, where $\sigma_{\nu}$ is the initial velocity dispersion of the atomic ensemble. The corresponding biases on the value of $g$ are therefore independent of $T$, and proportional to the temperature. In contrast, higher order terms, e.g spherical aberrations, give biases on $g$ which depend not only on the temperature but also on the value of $T$.

\begin{figure}[h]
\centering
\includegraphics[scale=1]{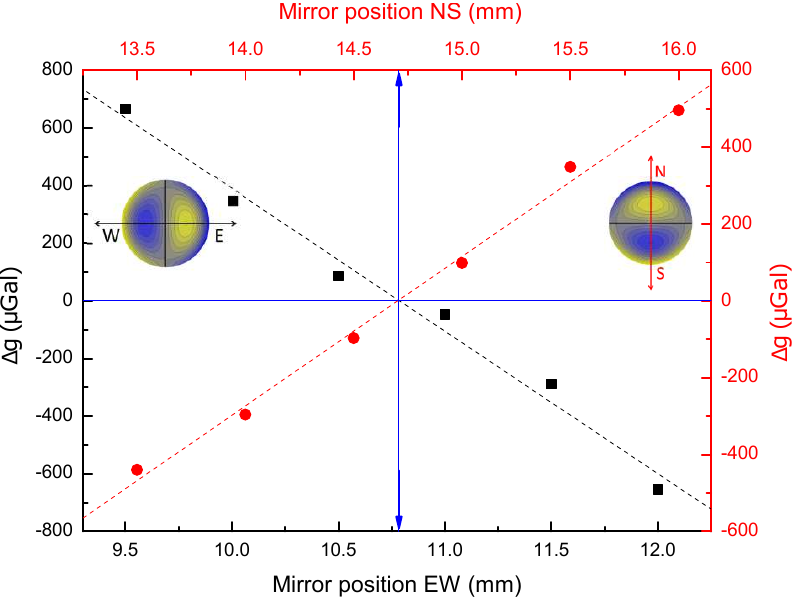}
\caption{Effect of comas, of $0.4\lambda$ amplitude, on the gravity measurement. The deformable mirror was displaced on the Est-West (resp. North-South) direction with a coma 0$\degree$ deformation (black squares) (resp. coma 90$\degree$ (red circles)).}
\label{fig:DMadjustment}
\end{figure}

\section{Measurements with an atom interferometer}

A good control of the mirror and atoms parameters (such as centering and alignments) is necessary to characterize the DM impact on the interferometer, and compare experimental results with a model of the experiment. For that purpose, a first coarse adjustment of the mirror position was initially performed by maximizing the number of detected atoms. Due to mechanical tolerance and alignment errors, the atomic cloud is not necessarily perfectly aligned with the Raman beams and the center of the detection area. Thanks to additional bias coils, the initial position of the cloud was set so as to maximize the contrast of the interferometer, which corresponds to placing the atomic cloud at the center of the Raman beams. Once the atomic cloud position is fixed, we took advantage of the property of the coma aberration to make a finer adjustment of the DM onto the center of the atomic cloud. Significant coma aberrations were applied on the DM, and the differential bias on $g$ was measured as a function of the mirror position, with respect to the FM configuration, as shown in Fig. \ref{fig:DMadjustment}. As expected, linear dependencies are observed through the East-West (EW) and North-South (NS) directions, which are aligned with the proper axes of the DM. The zero crossing positions (marked by the blue arrows) correspond to the best DM alignment with the center of the atomic cloud.

In order to determine the relationship between the wavefront and the interferometric phase shift, a simulation of an interferometer using a DM has been developed. This Monte-Carlo simulation reproduced the experiment described in Fig. \ref{fig:Setup and principle}, taking into account the parameters of the atomic source, the inhomogeneity of the detection response (as in \cite{Farah:14}) and the wavefront aberrations. To characterize the DM, the bias of different aberrations on $g$ were measured by varying the mirror shape for $T$=58~ms. For these measurements, the short term sensitivity was in the range 100-200~$\mu$Gal at 1~s. A summary of the comparison between the numerical simulations and the experimental results is shown in Table \ref{tab:SimulationsVsMeasure}. The bias of the focus and the spherical aberration on $g$ were measured for weak deformations of the DM. We found a good agreement for the spherical aberration, but a significant difference for the focus, which is not explained. To evaluate the effect of the comas, we set a fixed deformation of $0.4\lambda$ and displaced the DM on the EW (for the coma 0$\degree$) and NS (for the coma 90$\degree$) directions, for a fixed position of the cloud Fig. \ref{fig:DMadjustment}. Here also, experiments were in good agreement with the simulations.

\begin{table}[h]
	\centering
	\caption{Comparison between simulations and measurements of different aberration biases on $g$.}
	\begin{tabular}{ccc}
		\hline
		Aberration & Measurement & Simulation \\
		\hline
		Focus & 2991(55)~$\mu$Gal/$\mu$m & 3652(10)~$\mu$Gal/$\mu$m \\
		Spherical ab. & 3172(110)~$\mu$Gal/$\mu$m & 3275(10)~$\mu$Gal/$\mu$m \\
		Coma 0$\degree$/EW & -494(30)~$\mu$Gal/mm & -523(1)~$\mu$Gal/mm \\
		Coma 90$\degree$/NS & -503(14)~$\mu$Gal/mm & -522(1)~$\mu$Gal/mm \\
		\hline
	\end{tabular}
	\label{tab:SimulationsVsMeasure}
\end{table}

Given the measurement of the DM flatness in FM configuration, we now evaluate the corresponding bias on the gravity measurement. For that, we consider only the contributions having revolution symmetry such as the focus and the different orders of spherical aberrations. By weighting these contributions with their corresponding measured sensitivities, we estimate a relatively large bias on the gravity measurement of the order of 30~$\mu$Gal. This FM calibration is performed by the manufacturer by minimizing the global RMS error, which is not best suited for our application for which one would minimize aberrations of revolution symmetry (such as the focus, the spherical aberrations, ...) and would tolerate higher residuals on the other aberrations (tilts, astigmatisms, comas, trefoils, ...). Given the excellent resolution on the actuators displacement (of order of 50~pm in principle), lower biases could be obtained by adjusting the mirror with these constraints, which would improve the accuracy of the gravimeter. Alternatively, comas could be minimized in order to reduce the sensitivity to the initial position of the atomic cloud (see Table \ref{tab:PhaseShiftAberration} and the measurements below), which would improve the long term stability of the measurement.

\begin{figure}[h]
	\centering
	\begin{subfigure}[b]{0.23\textwidth}
		\includegraphics[width=\textwidth]{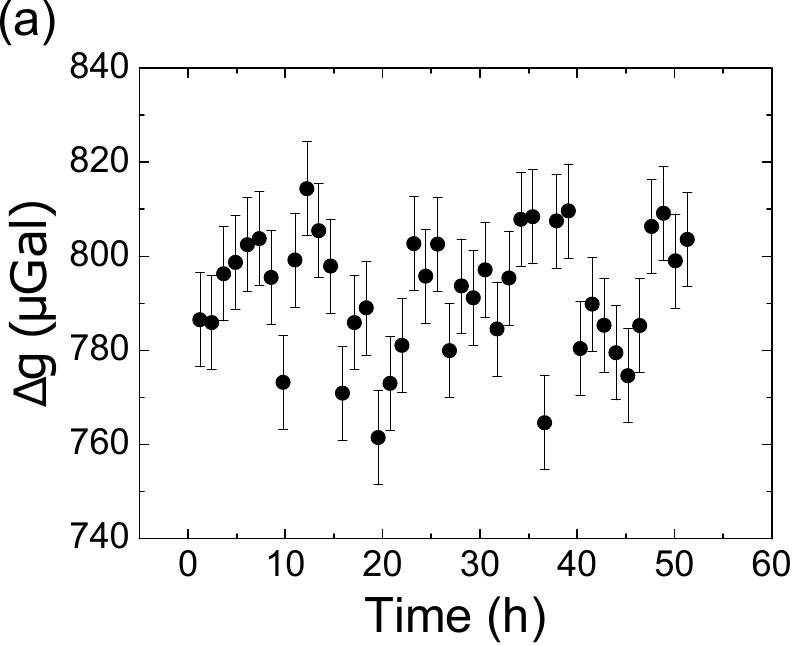}
	\end{subfigure}
	\begin{subfigure}[b]{0.23\textwidth}
		\includegraphics[width=\textwidth]{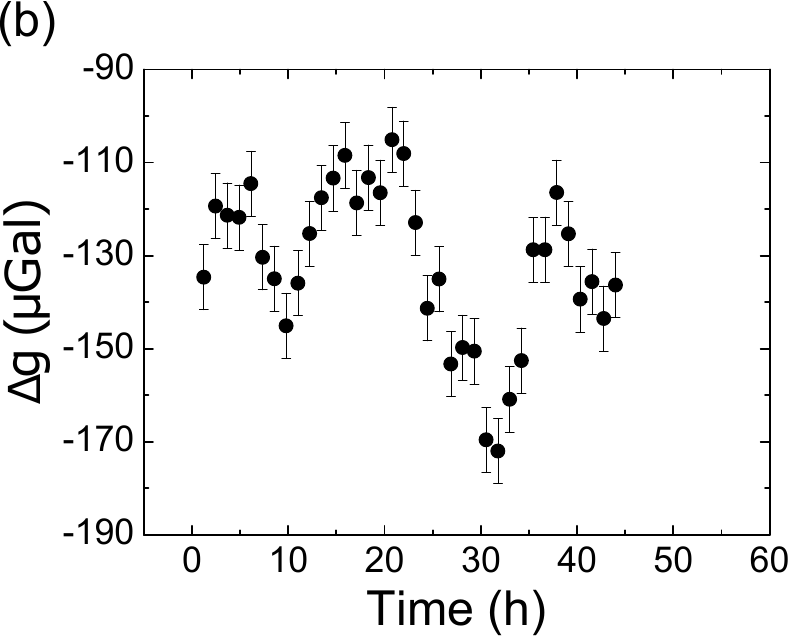}
	\end{subfigure}
	\caption{Differential gravity beetween (a) two different focus deformations of $0.1\lambda$ and  $0.6\lambda$ amplitudes and (b) a coma 90$\degree$ deformation with $0.1\lambda$ amplitude and the FM configuration.}
	\label{fig:AbMeas}
\end{figure}

We then evaluated the stability of the gravity measurements when deliberately applying selected aberrations using the DM. First, a differential measurement with two different amplitude of focus ($0.1\lambda$ and $0.6\lambda$) was performed over 2 days. Figure \ref{fig:AbMeas}(a) displays the results of this measurement, where each point is averaged over 4400~s (73~mn). The observed fluctuations around the average value of 792~$\mu$Gal are consistent with a white noise, as the corresponding Allan standard deviation is found to decrease as 1/$\sqrt{\tau}$ with the averaging time $\tau$. We reach a stability of 4~$\mu$Gal after 10~h, which corresponds to a remarkable relative stability of $0.5~\%$. This confirms the high stability of the DM in open loop. Then, we performed differential gravity measurements of a fixed coma 90$\degree$ with $0.1\lambda$ amplitude versus the FM configuration. We observed relatively large and well resolved variations, displayed on Fig. \ref{fig:AbMeas}(b), of the order of $\pm30~\mu$Gal over a day, which we attribute to slow fluctuations of the atomic source initial position (of the order of $\pm200~\mu$m) \cite{Farah:14}. These position fluctuations bias the gravity measurement in the presence of asymmetric wavefront distortions such as coma aberrations. 

\section{Compensation of wavefront aberrations}

Due to its high stability, the deformable mirror could in principle be used to correct the biases caused by the bottom window of the vacuum chamber and the quarter-wave plate as described in Fig. \ref{fig:Setup and principle} (Right). However, the optical flatness of the viewport was not measured before being installed. Furthermore it is very likely that its properties have been modified by its installation in the experimental chamber, due to mechanical and thermal stresses during the pumping process. 

As mentioned before, the wavefront aberration effect on the gravity signal depends on many experimental parameters, which allows to get some insight on their shape and amplitude. For instance, the size of the atomic detection \cite{Schkolnik:15} or the aperture of the Raman beams \cite{Zhou:16} acts as a filter for the atomic trajectories which contribute to the interferometer signal. Also, the effect of wavefront distortions gets modified when varying the temperature or modifying the initial position of the cloud (see table \ref{tab:PhaseShiftAberration}). Remarkably, increasing the initial size of the atomic cloud reduces the contribution of high frequency components of the wavefront \cite{Louchet:10}. Measurements of the interferometer phase versus the above mentioned parameters can be compared with phase shifts calculated for different models of these aberrations \cite{Louchet:10,Schkolnik:15,Zhou:16}. But, in the absence of an a priori knowledge of the wavefront, the deconvolution from the interferometer response and the averaging over the trajectories is a difficult task, due to non-unicity of the solution of the inverse problem \cite{Louchet:10}.
As a way to overcome this, spatially resolved detection, such as the point source interferometry imaging technique demonstrated in \cite{Kasevich:13}, allows for the measurement of the phase shift as a function of the transverse position in the interferometer laser beam, which is then related only to the initial transverse velocity. This renders the deconvolution simpler and offers the possibility of more accurately reconstructing the wavefront, and thus retrieving the resulting wavefront aberrations. In that case, one would be able to compensate for these distortions by using a DM. Unfortunately our sensor geometry is not well adapted for the implementation of this technique, due to lack of optical access.

\begin{figure}[h]
	\centering
	\includegraphics[scale=1]{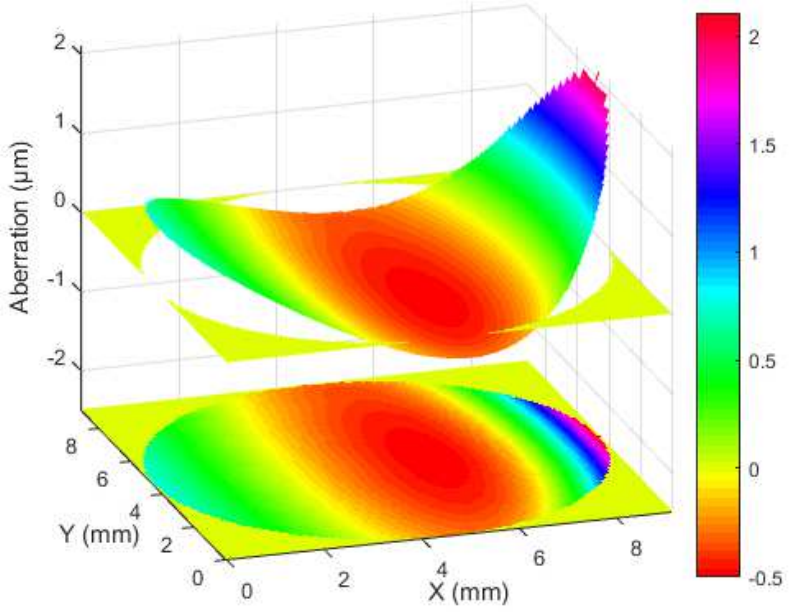} 
	\caption{Aberrations of the additional window measured by a Shack-Hartmann sensor in direct 	transmission through a 9~mm aperture diaphragm area.}
	\label{fig:WindowAberrations}
\end{figure}

\begin{figure}[h]
	\centering
	\includegraphics[scale=1]{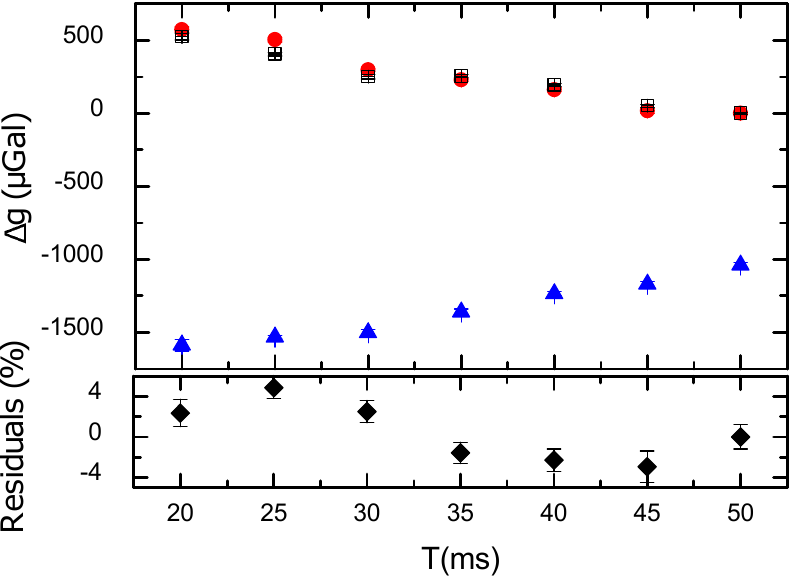} 
	\caption{Gravity measurements for different interferometer times $2T$ with and without additional window, with and without aberration correction. (Up): Open black squares represent the measurements realized with the deformable mirror (DM) in flat map configuration before inserting the additional window. Blue triangles, after inserting the window. Red circles display the measurements realized with the aberrations compensated by the DM. (Down): Relative residuals after the compensation.}
	\label{fig:DefectCompensation}
\end{figure}

Instead, and for a proof of principle, a well characterized optical element was inserted between the bottom window and the mirror, and its effect on the interferometric measurement was compensated by adapting the shape of the mirror. More precisely, we used an additional window of low optical quality, selecting a 9~mm diameter area which presented strong aberrations in order to generate a large bias on the measurements. The wavefront aberrations of this area were initially characterized using the SH, in direct transmission.
Figure \ref{fig:WindowAberrations} shows these aberrations which were decomposed on the Zernike polynomials basis. Dominant contributions were: 780(22)~nm of Astigmatism 0$\degree$, -480(15)~nm of Focus, -370(12)~nm of Coma 0$\degree$, and -60(6)~nm of Spherical Aberration. In order to compensate for the wavefront distortion caused by the additional window, the DM was shaped following the same aberration by summing the above contributions with their respective amplitudes. To assess the efficiency of the wavefront correction by the DM, a set of gravity measurements has been realized before and after the installation of the additional window for several interferometer time $2T$. Each gravity value is obtained by averaging the results of two measurements performed with two opposite wavevector directions in order to reject most of the systematic effects \cite{Louchet:10}. The gravity measurements reported below are differential taking the value of $g$ measured for $T=$ 50 ms with the DM in FM configuration as a reference. For the reference measurements, the contrast of the interferometer is 17$\%$. The measurement process was done in four steps, and the results are displayed in Fig. \ref{fig:DefectCompensation}.

First, a series of reference measurements (represented by open squares) were realized with the DM in FM configuration before adding the window. Note that in all the differential measurements, the systematic effect due to the two-photon light shift \cite{Gauguet:08} was not corrected for, which explains most of the observed variation of the measured values of $g$ as a function of $T$.
Second, we added the window, we observed a reduction of the contrast down to 10$\%$ and we measured a change of the gravity value as large as -1040(10)~$\mu$Gal for $T=50$ ms with respect to the reference configuration, keeping the DM in FM configuration. 
Using Table \ref{tab:SimulationsVsMeasure}, we expected a variation of -1626(99)~$\mu$Gal of gravity due to the effect of the window aberrations. We attribute the difference between the calculated and measured values to the DM nonlinearities, which are significant for high deformations ($>0.3\lambda$).
Third, we repeated the gravity measurements for different values of $T$, represented by the blue triangles, in the presence of the additional window keeping the DM in FM configuration.

Finally, the DM was shaped according to the corrections described earlier and we recovered the initial contrast of 17$\%$ at $T=$ 50 ms, which we take as a first evidence of the efficiency of the wavefront correction. This is confirmed by a last series of differential measurements, displayed as red circles. We find a good agreement with the initial measurements performed without the additional window, which demonstrates the efficiency of the compensation. Relative residuals (displayed as black diamonds at the bottom of Fig. \ref{fig:DefectCompensation}) lie in between $\pm4\%$. These differences can be explained by residual imperfections of the correction and fluctuations of the two-photon light shift.
 
\section{Conclusion}

In conclusion, we have demonstrated that the use of an appropriate deformable mirror allows to correct the wavefront aberrations in atomic interferometers. Though the compensation has been demonstrated here for the large distortions induced by an additional window of poor optical quality, it should also be effective for weaker aberrations thanks to the high resolution of the actuation and the excellent stability of the mirror. This could be demonstrated in state of the art atom gravimeters, such as those of \cite{Louchet:10,freier:16,Hu:13}.

In addition, the large dynamical range of the DM and its short response time would enable, at the same time, to suppress Coriolis acceleration (compensating Earth rotation by counter-rotating the mirror during the interferometer sequence \cite{Muller:12}) and reject ground vibration noise (by translating the mirror surface in real time \cite{Chu:99} or right before the last Raman pulse, similar to \cite{Lautier:14}). These compensation techniques can be extended to other instruments based on atom interferometry, such as gravity gradiometers and gyroscopes. In particular, they would be relevant for large scale experiments, such as based on large momentum transfer beam splitters and/or long interferometer times. Indeed, in these experiments, the effect of wavefront aberrations scales as the effective momentum $n\hbar k$ imparted to the atoms, and the effect of high order aberrations onto the inertial measurement increases with the interferometer duration $2T$.

\begin{acknowledgments}
This work is supported by CNES (R\&T R-S11/SU-0001-030). ML thanks Muquans for financial support. The authors thank Boston Micromachines Corporation for their assistance, A. Landragin for fruitful discussions, and E. Cocher, C. Janvier, H. Bouchiba and M. Mazet for their contributions in the early stage of the experiment.
\end{acknowledgments}

\bibliography{Bibliographie}

\end{document}